\documentclass[12pt]{article}
\usepackage{times}
\usepackage{graphicx}
\usepackage{booktabs}
\usepackage{amsmath}
\usepackage{natbib}
\usepackage{url}
\title{You Talkin' to Me?: A Network Analysis of Gendered Speaker-Addressee Patterns in Film Screenplays}
\author{Samin Khan$^{1}$, Camilla Griffiths$^{1}$, Shrikanth Narayanan$^{2}$\\
Dan Jurafsky$^{1}$, Sabyasachee Baruah$^{2}$\\
\texttt{samink@stanford.edu}\\
$^{1}$Stanford University\\
$^{2}$University of Southern California}
\date{}
\begin{document}
\maketitle
    
\begin{abstract}
\noindent \textbf{Objective:} This paper investigates the gendered structure of speaker-addressee relationships in film dialogue, asking not merely who speaks, but who is spoken to and how conversational dynamics unfold across gender lines. \\
\textbf{Methods:} Using a manually annotated dataset of 4,600 directed dialogue events from 38 film screenplays, we apply network analysis, chi-squared tests, paired statistical comparisons, and participation shift (p-shift) analysis across three studies. \\
\textbf{Key Findings:} Male characters dominate as both speakers and addressees corpus-wide, even in scenes with more women; cross-gender dialogue is directionally symmetric on average but clustered at the film level; and same-gender (homophilous) turns diffuse conversational attention while cross-gender (heterophilous) turns produce tighter dyadic reciprocation. \\
\textbf{Conclusion:} Gender bias in film dialogue operates through the architecture of conversation itself—through exclusion from interaction and structural positioning as addressees—rather than within-conversation directional imbalance alone.
\end{abstract}
    
    \section{Introduction}
    Films both reflect and shape social norms, making the study of representation within them a critical endeavor. In recent years, computational methods have enabled large-scale analyses of gender in film, often focusing on a demographic group's total screen time or the linguistic frames used to portray them, such as power and agency \citep{sap2017,martinez2022,bamman2024}. While this work has provided invaluable insights, it often overlooks a fundamental component of narrative: the relational structure of character interaction. Beyond simply asking who is on screen, or what they are saying, we ask: who gets to speak to whom?
    
    This paper addresses a key gap in the literature by focusing on the structure of dialogue itself. The architecture of conversation—who initiates, who responds, and who is brought into a discussion—is a primary mechanism through which narrative develops, information is revealed, and character importance is established. However, identifying the specific addressee in scenes with three or more characters has proven to be a significant challenge for automated computational models \citep{agarwal2015,westphal2018}. Our work overcomes this limitation by using a manually annotated dataset of speaker-addressee pairs from 38 film screenplays.
    
    Using this rich relational data, we employ network analysis methods to uncover 
    the gendered patterns that govern character interactions. The paper follows a 
    three-step narrative. First, we ask whether gender predicts the basic direction of 
    conversational attention: who addresses whom, and whether women become stronger 
    conversational targets when women are more present. Second, because aggregate 
    address patterns could be produced either by imbalance within cross-gender 
    conversations or by exclusion from them, we test whether male-to-female and 
    female-to-male speech are directionally symmetric when cross-gender exchange occurs. 
    Third, if cross-gender speech is roughly symmetric, we ask what mechanism might still 
    reproduce male-centered interactional structure: whether different gender pairings 
    alter the downstream flow of conversation by closing into dyads or spreading to 
    additional participants. These levels motivate the following questions:

    \begin{itemize}
        \item \textbf{RQ1: Does speaker gender predict addressee gender, and does increasing female 
    presence alter the distribution of conversational targets?} 
        \item     \textbf{RQ2: When male and female characters speak across gender lines, is speech 
    directionally symmetric, or does one gender systematically direct more speech 
    toward the other?} 
        \item     \textbf{RQ3: Do homophilous and heterophilous initial turns produce systematically 
    different distributions of subsequent participation shifts?} 
    \end{itemize}

    \section{Related Work}
    This research is situated at the intersection of three domains: gender and media representation, narrative network analysis, and the computational study of dialogue.
    
    Prior work on gender in film has revealed significant disparities in representation. Feminist film theory has long argued that classical narrative cinema organizes looking, agency, and story movement around masculine subject positions, positioning women more often as objects of attention than as agents who structure the narrative field \citep{mulvey1975,delauretis1987,hooks1992}. Quantitative work has extended these concerns by measuring disparities in screen time and speaking time \citep{lauzen2025,smith2020,guha2015}, as well as subtler linguistic cues. For instance, some scholars have found that female characters are often framed with less agency through verb choice \citep{sap2017}, while others have identified systematic gender differences in the actions assigned to characters in movie scripts \citep{martinez2022}. These studies highlight important patterns of representation but tend to focus on the attributes of individual characters rather than the relational system in which they are embedded.
    
    A second stream of research has applied network science to narrative texts, extracting social networks from screenplays and literary works to model character relationships \citep{elson2010,bost2019,kagan2019}. Foundational work in this area has provided tools for extracting interaction networks and measuring character centrality \citep{elson2010,jones2020}. Some have argued that while female characters may have substantial screen time, this does not guarantee narrative centrality \citep{jones2018}. The Bechdel-Wallace test and its computational descendants make a related relational claim: representation depends not only on whether women appear, but on whether women have interactional space with one another \citep{bechdel1985,agarwal2015,selisker2015}. However, these studies do not always explicitly examine gender as a primary variable in directed addressee networks, nor do they fully interrogate how structural positions and network-building roles within the narrative differ between male and female characters \citep{leontyeva2024}.
    
    Finally, our work is informed by computational linguistics, particularly the study of dialogue. While sophisticated models exist for analyzing two-party conversations \citep{sacks1974}, multi-party dialogue remains a challenge \citep{stivers2021}. The ambiguity of "who is talking to whom" in a group setting has been a barrier to large-scale, automated analysis of narrative interaction \citep{agarwal2015,westphal2018}. Our project addresses this directly through manual annotation, providing the clean, directed interaction data necessary for a precise network analysis.
    
    By synthesizing these areas, this paper offers a novel contribution. While others have looked at gender in film and some have analyzed narrative networks, our work sits at the intersection, using a robust, manually-coded dataset to apply network methods specifically to the investigation of gender dynamics in a way that moves beyond simple counts to the very structure of the story.
    
    \section{Data}
    
    The dataset for this study was constructed from 38 film screenplays sourced from the Internet Movie Script Database (IMSDB). The 38 films comprise all final-draft or post-production scripts available on IMSDB, selected to ensure alignment between the annotated screenplay text and the dialogue as filmed. The selection spans four decades from \textit{Last Tango in Paris} (1972) to \textit{Logan} (2017). In total, our analysis is based on approximately 4,600 lines of dialogue where a direct addressee was manually annotated, forming a robust dataset for mapping interactional networks.
    
    The annotation process was performed by two undergraduate psychology students. For each line of dialogue in a screenplay, the annotators identified the speaker and, where possible, the specific character being addressed. Scenes with three or more characters present a significant challenge for automated addressee identification \citep{agarwal2015}; our manual annotation process was designed specifically to overcome this ambiguity and capture the directed nature of conversational exchanges. The annotators achieved an inter-rater reliability of over 80\%, ensuring the consistency of the addressee data.
    
    Character gender was inferred from character first names using the \texttt{gender-guesser} Python package~\citep{genderguesser2016}, a free and open-source tool that has been applied in prior computational social science studies~\citep{santamaria2018}. This procedure approximates perceived binary gender in the screenplay corpus rather than lived gender identity. It therefore cannot capture nonbinary, transgender, or otherwise non-cisgender identities, and it should not be treated as a general model of gender. For these particular films, which are Western productions with predominantly Western character names and conventionally gendered character presentation, the binary approximation is a pragmatic but limited operationalization for studying large-scale narrative patterns. A known additional limitation is that name-based tools may misclassify non-Western names \citep{santamaria2018,boyce2024}.
    
    The final dataset is structured as a relational event list, where each row corresponds to a single instance of a speaker addressing an addressee. The key columns for our analysis include \texttt{filename} (the film title), \texttt{curr\_speaker} (the character speaking), \texttt{addressee} (the character being spoken to), \texttt{curr\_gender} (gender of the speaker), and \texttt{addressee\_gender} (gender of the addressee). This granular, annotated data allows for a precise analysis of the directed interaction networks within each film's narrative \citep{bost2019}.
    
 \section{Study 1: Gendered Addressing Patterns in All Scenes vs. Female-Dominant Scenes}
\subsection{Research Question and Motivation}
Prior work on gender in film has largely focused on gender-based measures of presence, such as screen time and speaking time \citep{lauzen2025,smith2020}, or on linguistic framing, such as agency and power expressed through language \citep{sap2017,martinez2022}. These studies do not capture the gender-based directionality of dialogue by analyzing who is addressed by whom. This study addresses that gap by answering the question: Does speaker gender predict addressee gender?
Work in narrative network analysis has examined character centrality and interaction structure \citep{elson2010,jones2020,bost2019}, but has typically not treated gender as a primary variable in directed interaction patterns. As a result, an important question remains underexplored: when women are present in greater numbers within a scene, does conversational attention shift toward them, or do male characters remain the primary recipients of speech \citep{kagan2019,jones2018}?
To address this gap, we ask: \textbf{Does speaker gender predict addressee gender, and does increasing 
female presence alter the distribution of conversational targets? (RQ1)} 
\subsection{Methods}
Our unit of analysis is the directed speaker-addressee dyad. Each interaction is categorized into one of four types: male-to-male (M$\rightarrow$M), male-to-female (M$\rightarrow$F), female-to-male (F$\rightarrow$M), and female-to-female (F$\rightarrow$F). We performed two separate analyses: one on the entire dataset of interactions and another on the subset of interactions occurring only in female-dominant scenes.
For both datasets, we constructed a 2$\times$2 contingency table of speaker gender versus addressee gender. We then employed Pearson's chi-squared test of independence to determine whether an association exists between the gender of the speaker and the gender of the addressee. The null hypothesis ($H_0$) is that these two variables are independent.
\subsection{Results}
The analysis reveals a significant association between speaker and addressee gender in both conditions, but the nature of this association changes notably when moving from all scenes to female-dominant scenes.
\subsubsection{All Scenes}
Across the entire corpus, dialogue is dominated by male-to-male interactions. As shown in Table~\ref{tab:all_scenes}, M$\rightarrow$M dialogue accounts for approximately 40.2\% of all addressed speech. A chi-squared test confirms a highly significant relationship between speaker and addressee gender ($\chi^2(1, N=2532)=97.70, p<.001$; specifically, $p=4.87 \times 10^{-23}$). This result is consistent with prior quantitative studies showing male overrepresentation in screen presence and narrative participation \citep{guha2015,bamman2024}, but it extends those findings from presence to directed conversational attention: men are not only more visible or more often speaking, they are also the dominant recipients of speech.
\begin{table}[h!]
\centering
\caption{Interaction Counts and Proportions: All Scenes}
\label{tab:all_scenes}
\begin{tabular}{@{}lrr@{}}
\toprule
\textbf{Gender Interaction} & \textbf{Count} & \textbf{Proportion} \\ \midrule
male$\rightarrow$male & 1019 & 0.402 \\
male$\rightarrow$female & 677 & 0.267 \\
female$\rightarrow$male & 667 & 0.263 \\
female$\rightarrow$female & 169 & 0.067 \\ \bottomrule
\end{tabular}
\end{table}
\subsubsection{Female-Dominant Scenes}
In scenes where female speakers are at least as numerous as male speakers, the interaction dynamics shift substantially. The chi-squared test remains highly significant, indicating that gender continues to structure dialogue ($\chi^2(1, N=1205)=652.22, p<.001$; specifically, $p=7.37 \times 10^{-144}$).
However, as detailed in Table~\ref{tab:fem_dom}, the dominant interaction is no longer male-to-male, which drops to just 1.6\% of dialogue. The most frequent interaction type becomes female-to-male, accounting for 44.6\% of all dialogue in these scenes, closely followed by male-to-female at 41.2\%. Female-to-female interaction rises from 6.7\% corpus-wide to 12.7\% in female-dominant scenes; a two-proportion test confirms that this increase is statistically reliable ($z=6.13$, $p=8.65 \times 10^{-10}$). Yet the substantive pattern remains strongly cross-gender: even in scenes where women are at least as numerous as men, female characters address men far more often than they address other women (44.6\% vs. 12.7\%).
\begin{table}[h!]
\centering
\caption{Interaction Counts and Proportions: Female-Dominant Scenes}
\label{tab:fem_dom}
\begin{tabular}{@{}lrr@{}}
\toprule
\textbf{Gender Interaction} & \textbf{Count} & \textbf{Proportion} \\ \midrule
female$\rightarrow$male & 537 & 0.446 \\
male$\rightarrow$female & 496 & 0.412 \\
female$\rightarrow$female & 153 & 0.127 \\
male$\rightarrow$male & 19 & 0.016 \\ \bottomrule
\end{tabular}
\end{table}
\subsection{Discussion}
The findings from Study 1 reveal a robust and previously underexamined pattern in gendered dialogue. Consistent with prior work on homophily in social networks \citep{mcpherson2001} and narrative interaction structures \citep{elson2010}, the analysis of all scenes confirms a strong tendency toward male homophily, positioning male characters within dense, self-referential interaction structures. This aligns with prior findings that centrality in narrative networks is not evenly distributed across characters \citep{jones2020,kagan2019}.
However, the analysis of female-dominant scenes challenges a key implicit assumption in the literature: that increasing the presence of women leads to increased interaction among women \citep{leontyeva2024}. While the numerical presence of female characters substantially reduces male-to-male dialogue, it does not produce a corresponding rise in female-to-female interaction. Instead, the dominant pattern becomes cross-gender interaction, particularly female-to-male dialogue.
This result highlights a limitation of existing representation metrics. Measures such as screen time and speaking time \citep{smith2020,lauzen2025}, as well as large-scale linguistic analyses of gendered portrayal \citep{sap2017,martinez2022}, primarily focus on attributes of individual characters rather than the relational structure of interaction. As a result, they implicitly assume that presence or participation translates into relational influence. Our findings suggest otherwise. Even when women are numerically well represented within a scene, they do not become the primary targets of speech. Instead, conversational attention remains disproportionately directed toward male characters.
From a network perspective, this suggests that male characters occupy a structurally privileged role as addressees, functioning as central recipients of dialogue even in contexts where they are not numerically dominant. This distinction between speaking and being addressed introduces an additional dimension of narrative centrality that is not captured by existing approaches. Prior work has shown that structural position within a network can differ from surface-level participation \citep{jones2018,leontyeva2024}, and our results extend this insight by demonstrating that conversational targeting is itself systematically gendered.
More broadly, Study 1 demonstrates that representation cannot be fully understood through counts of characters or lines of dialogue alone. The directionality of interaction—who speaks to whom—constitutes a critical and underexplored dimension of narrative structure \citep{selisker2015}. By focusing on conversational targets, this study contributes a structural perspective to the analysis of gender in media, revealing a deeper bias in which male characters remain central as recipients of speech even when women are numerically well represented.
\section{Study 2: Directionality of Cross-Gender Dialogue}
\subsection{Research Question and Motivation}
Study 1 suggests that gender structures the targets of speech in aggregate. Yet the asymmetry of cross-gender dialogue itself remains an open question. \textbf{When male and female characters engage in cross-gender dialogue, 
is speech exchanged with directional symmetry, or does one gender 
systematically direct more speech toward the other? (RQ2)}
\\ \\
This question is distinct from the proportion of interactions involving each gender. Even in a corpus where M$\rightarrow$F and F$\rightarrow$M interactions appear balanced in aggregate, directional asymmetry may manifest at the scene level—driven by genre conventions, narrative context, or the relational roles assigned to characters \citep{martinez2022,jones2018}. By measuring the directional share of cross-gender speech within individual scenes, we test whether male-to-female speech systematically outnumbers female-to-male speech, or whether the two are exchanged with rough symmetry.

\subsection{Methods}
For each mixed-gender scene, we count the number of male-to-female and female-to-male directed utterances. We then compute the scene-level male-to-female share:
\[
\text{M}\rightarrow\text{F share}_s =
\frac{\#(\text{M}\rightarrow\text{F})_s}{\#(\text{M}\rightarrow\text{F})_s + \#(\text{F}\rightarrow\text{M})_s}.
\]
Scenes with no cross-gender dialogue are excluded from this study because directionality is undefined. We test whether the scene-level share differs from 0.5 using both a paired $t$-test on the difference between M$\rightarrow$F and F$\rightarrow$M counts and a Wilcoxon signed-rank test as a nonparametric robustness check. We also classify scenes as male-to-female dominant, female-to-male dominant, or balanced using thresholds of $>0.67$, $<0.33$, and the interval between them, respectively. Finally, to make the scene-level heterogeneity interpretable, we select three representative network visualizations from high-signal scenes with at least six cross-gender utterances.

\subsection{Results}
\subsubsection{Aggregate Symmetry}
Across the 166 mixed-gender scenes, the corpus contains 677 M$\rightarrow$F utterances and 667 F$\rightarrow$M utterances, yielding an overall M$\rightarrow$F share of 0.504. The mean scene-level share is 0.491 (median = 0.500). Neither the paired $t$-test ($t(165) = 0.26$, $p = .798$) nor the Wilcoxon signed-rank test ($W = 4106$, $p = .836$) yields a significant result. We therefore fail to reject the null hypothesis of directional symmetry at the aggregate level.
\begin{table}[h!]
\centering
\caption{Summary of Cross-Gender Dialogue Directionality}
\begin{tabular}{@{}lcc@{}}
\toprule
\textbf{Measure} & \textbf{Value} \\
\midrule
Mixed-gender scenes & 166 \\
Films & 27 \\
Total M$\rightarrow$F utterances & 677 \\
Total F$\rightarrow$M utterances & 667 \\
Overall M$\rightarrow$F share & 0.504 \\
Mean scene M$\rightarrow$F share & 0.491 \\
Median scene M$\rightarrow$F share & 0.500 \\
Paired $t$-test ($p$) & .798 \\
Wilcoxon signed-rank ($p$) & .836 \\
\bottomrule
\end{tabular}
\end{table}
\subsubsection{Scene-Level Variation}
Despite aggregate symmetry, the distribution of $M \rightarrow F$ share reveals substantial scene-level heterogeneity. Using thresholds of $< 0.33$ and $> 0.67$ to define directionally skewed scenes, we find that 28 scenes (16.9\%) are male-to-female dominant, 31 scenes (18.7\%) are female-to-male dominant, and the majority—107 scenes (64.5\%)—exhibit balanced cross-gender exchange.
Directional skew is also organized at the film level. Films such as \textit{Wild Hogs} (2007; mean M$\rightarrow$F share = 1.00), \textit{Fright Night} (1985; 0.63), and \textit{Logan} (2017; 0.62) are consistently male-directed in their cross-gender speech, whereas \textit{Red Riding Hood} (2011; 0.25), \textit{Cherry Falls} (2000; 0.28), and \textit{Legion} (2010; 0.33) are systematically female-directed. This film-level clustering suggests that directional patterns are not random noise but reflect structural narrative conventions tied to genre and casting \citep{kagan2019}.
\subsubsection{Representative Scene Networks}
\begin{figure}[h!]
\centering
\includegraphics[width=0.78\textwidth]{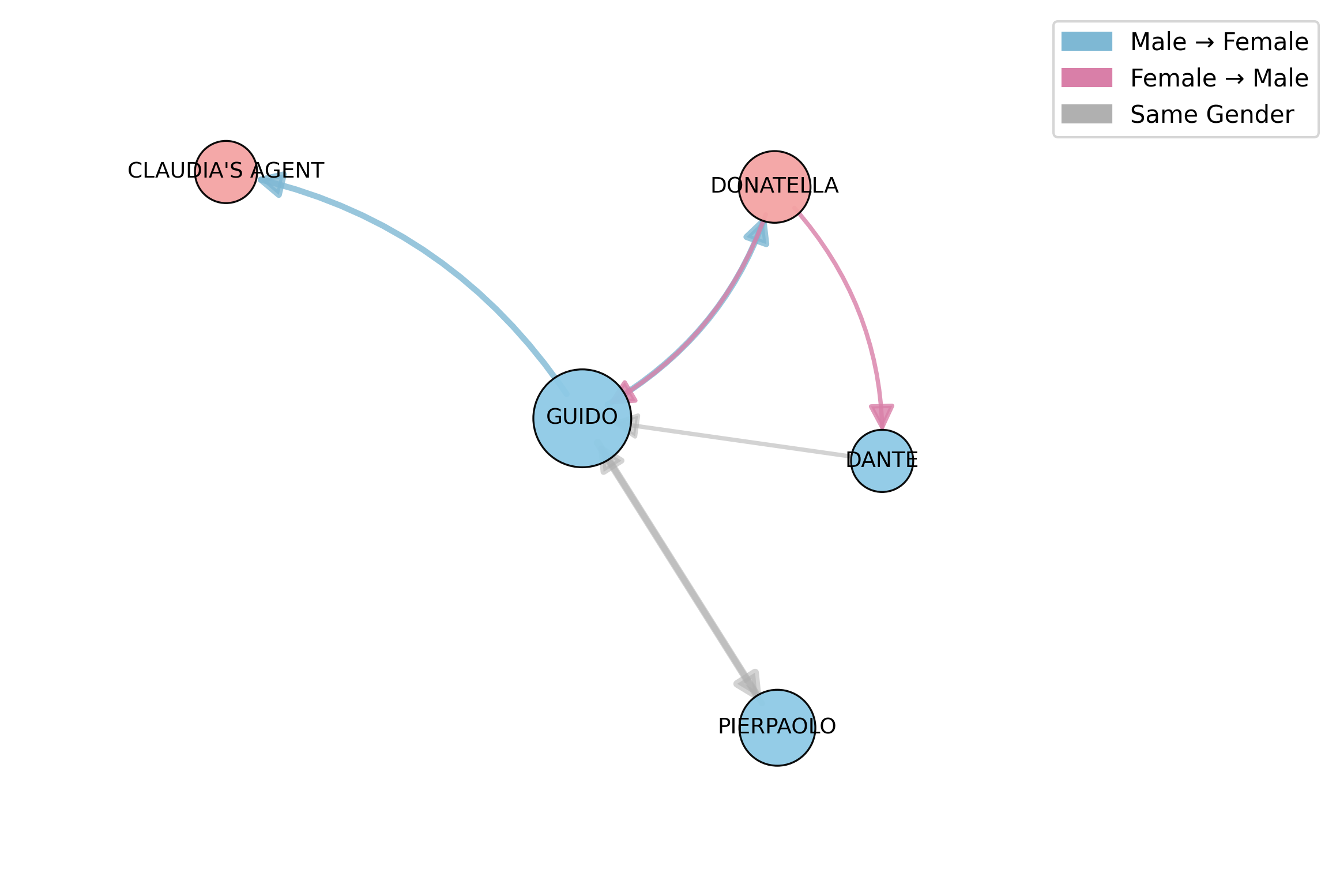}
\caption{This is an example scene from the male-to-female-skew cluster of film scenes, in which speaker-addressee patterns skew toward male characters directing more cross-gender speech to female characters. In this scene, Guido directs more cross-gender speech toward female characters than they direct toward him. Node size reflects total interaction volume, and arrow width reflects the number of directed utterances between characters. Scene 55 from \textit{Nine} [2009].}
\label{fig:nine_scene}
\end{figure}

\begin{figure}[h!]
\centering
\includegraphics[width=0.78\textwidth]{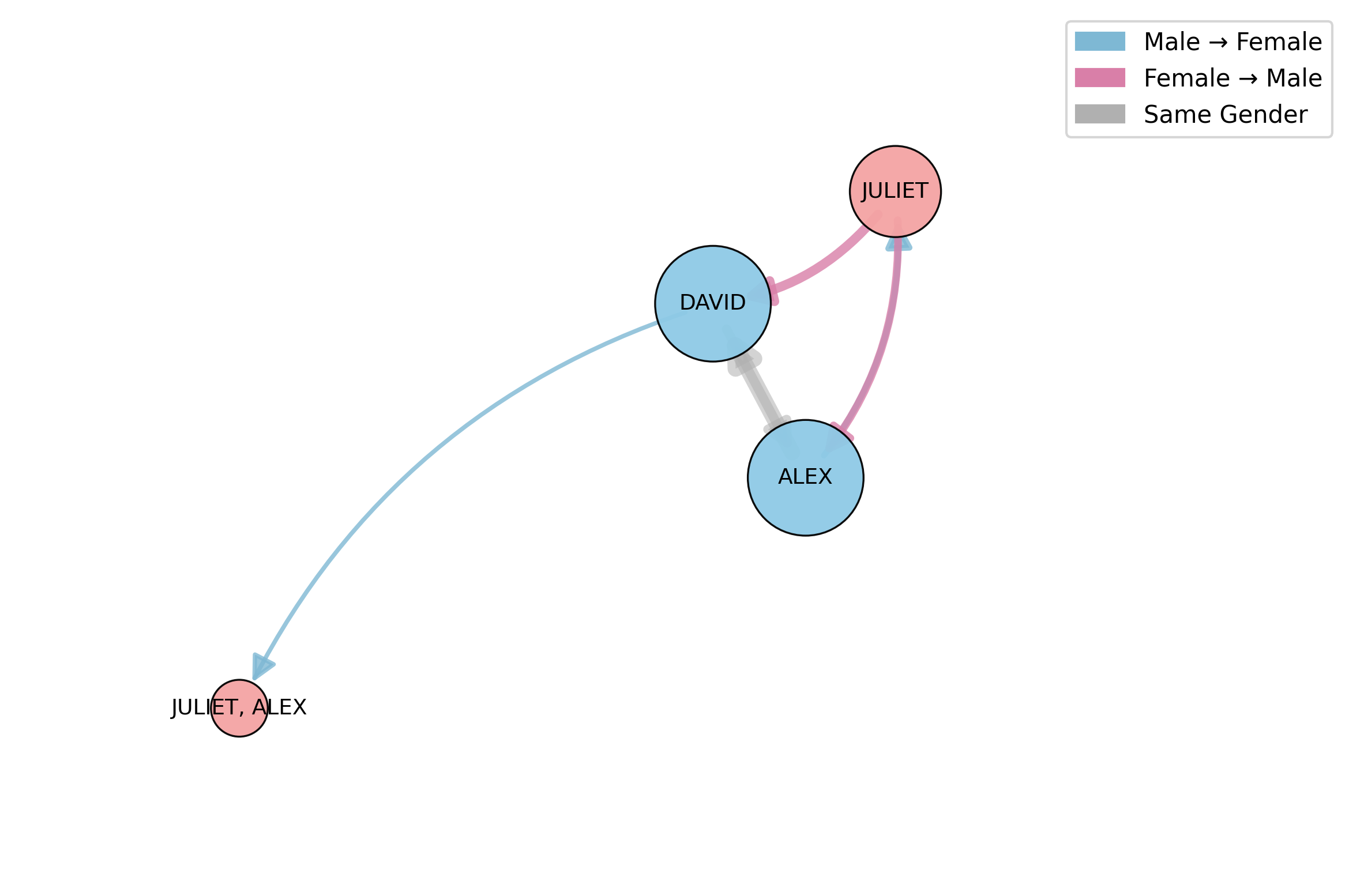}
\caption{This is an example scene from the female-to-male-skew cluster of film scenes, in which speaker-addressee patterns skew toward female characters directing more cross-gender speech to male characters. In this scene, Juliet directs more cross-gender speech toward David and Alex than male characters direct toward her. Node size reflects total interaction volume, and arrow width reflects the number of directed utterances between characters. Scene 100 from \textit{Shallow Grave} [1994].}
\label{fig:shallow_scene}
\end{figure}

\begin{figure}[h!]
\centering
\includegraphics[width=0.78\textwidth]{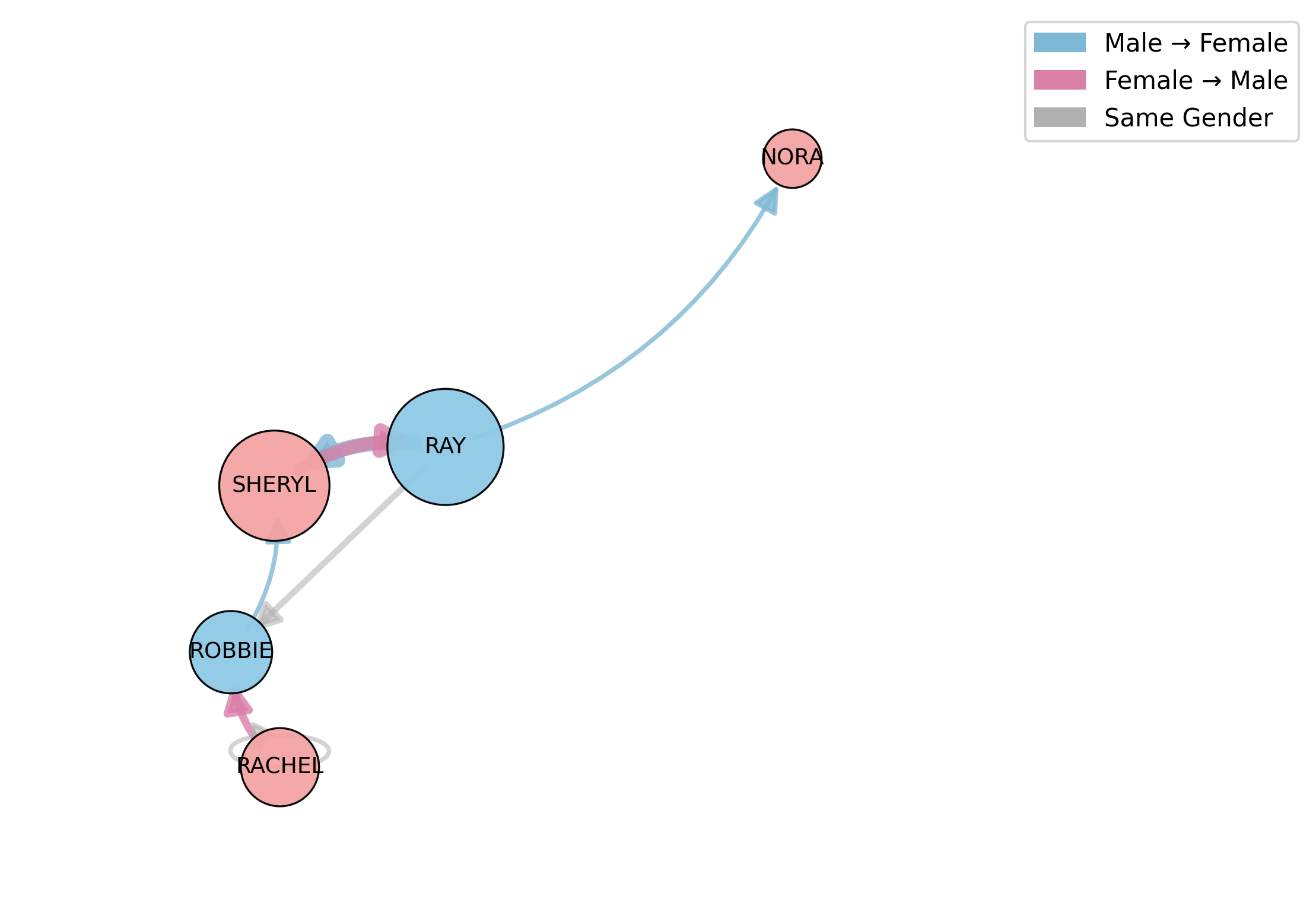}
\caption{This is an example scene from the balanced cross-gender cluster of film scenes, in which speaker-addressee patterns show roughly similar levels of male-to-female and female-to-male speech. In this scene, cross-gender speech is distributed across both directions rather than producing a clear directional skew. Node size reflects total interaction volume, and arrow width reflects the number of directed utterances between characters. Scene 29 from \textit{War of the Worlds} [2005].}
\label{fig:war_scene}
\end{figure}

Figures~\ref{fig:nine_scene}--\ref{fig:war_scene} illustrate the three scene-level patterns implied by the directionality analysis: male-to-female skew, female-to-male skew, and balanced exchange. The figures are therefore not additional tests; they make visible why an aggregate null result can coexist with meaningful local structure.
\subsection{Discussion}
The null result at the aggregate level is itself a finding of interest. The absence of a significant directional asymmetry in cross-gender dialogue contrasts with the strong male-centric patterns observed in Study 1, where M$\rightarrow$M dialogue dominated the overall distribution and female characters were disproportionately absent as speakers \citep{lauzen2025,bamman2024}. The present analysis suggests that \textit{when} cross-gender interaction does occur, it proceeds with rough reciprocity: female characters are neither systematically relegated to the role of addressee nor of initiator in mixed-gender exchanges.
This finding refines the picture from Study 1. The overall addressee disparity observed there is not primarily driven by male characters directing more speech at female characters than the reverse; rather, it is driven by the relative scarcity of cross-gender interaction itself compared to the volume of male-to-male dialogue. Put differently, the structural marginalization of female characters in film dialogue operates through \textit{exclusion} from the conversation--through the dominance of all-male interaction patterns--rather than through within-conversation directional imbalance. This exclusion-based interpretation connects the directed addressee results to feminist accounts of narrative cinema in which men more often occupy the active positions through which story movement and attention are organized \citep{mulvey1975,delauretis1987}, and to Bechdel-style critiques that foreground the scarcity of interactional space among women \citep{bechdel1985,selisker2015,agarwal2015}.
At the scene level, however, directional patterns are clearly non-random. The film-level clustering of skew values, with some films consistently male-directed and others consistently female-directed, points to genre-level conventions as a likely moderating factor. Action and ensemble-male films (e.g., \textit{Wild Hogs} (2007), \textit{Logan} (2017)) tend toward M$\rightarrow$F dominance, while horror and female-protagonist films (e.g., \textit{Cherry Falls} (2000), \textit{Red Riding Hood} (2011)) exhibit F$\rightarrow$M skew \citep{kagan2019}. This genre-level variation merits further investigation, as it suggests that the relational structure of gendered dialogue is not a uniform property of narrative film but is shaped by the gendered conventions of specific story types. The genders of the screenwriting staff may also matter: future work could test whether the gender composition of the screenwriters is associated with gender patterns amongst speakers and addressees in the screenplay. Studies 1 and 2 suggest that gendered differences in conversational centrality are not primarily explained by directional imbalance within cross gender dialogue. The relevant mechanism may lie in how different gender pairings shape the subsequent flow of conversation. Some exchanges may remain confined to a reciprocal dyad, while others may redirect attention or draw additional participants into the interaction. Study 3 tests this possibility by examining whether homophilous and heterophilous turns produce systematically different participation shifts.
\section{Study 3: Conversational Turn-Taking and Participation Shifts}
\subsection{Research Question and Motivation}
Study 1 established that gender structures \textit{who} speaks to \textit{whom} in aggregate, while Study 2 showed that cross-gender dialogue is not directionally imbalanced once it occurs. This combination creates a mechanism question. If male centrality is not mainly produced by one-way male-to-female dominance inside mixed-gender conversations, it may instead be produced by what different kinds of turns do next: some turns may keep attention locked in a dyad, while others may open the floor and pull additional characters into the exchange.
Prior work on conversation analysis has characterized these dynamics through \textit{participation shifts} (p-shifts), a taxonomy for describing how speaker-addressee configurations change from one turn to the next~\citep{gibson2003}. This framework builds on foundational work on turn-taking systems in natural conversation \citep{sacks1974,stivers2021}. Applying p-shift analysis to gendered interaction data allows us to test whether same-gender (homophilous) and cross-gender (heterophilous) turns differ not just in frequency but in their structural consequences for conversational flow \citep{hu2021}. Because homophilous turns in this corpus are predominantly male-to-male, this test directly evaluates a potential mechanism for the male-centered pattern in Study 1: whether male homophily is conversation-spreading while cross-gender exchange is conversation-closing.
We ask: \textbf{Do homophilous and heterophilous initial turns produce 
systematically different distributions of subsequent participation shifts, 
and are these differences robust to group size? (RQ3)}
\subsection{Methods}
\subsubsection{P-Shift Taxonomy}
We represent conversational flow as ordered two-turn sequences $(AB \rightarrow CD)$, where $A$ is the speaker of the first turn addressing $B$, and $C$ is the speaker of the next turn addressing $D$. We classify each transition into one of five p-shift types following \citet{gibson2003}:
\begin{itemize}
    \item \textbf{AB-AB}: continuation — the same speaker addresses the same addressee.
    \item \textbf{AB-BA}: reciprocation — the addressee turns back to speak to the original speaker.
    \item \textbf{AB-AC}: speaker readdress — the original speaker addresses a different participant.
    \item \textbf{AB-CA}: third-party entry — a new participant enters and addresses the original speaker.
    \item \textbf{AB-CD}: parallel shift — both speaker and addressee change to characters not in the prior turn.
\end{itemize}
Each initial turn $(A \rightarrow B)$ is labeled \textit{heterophilous} if speaker and addressee differ in gender, and \textit{homophilous} if they share gender.
\subsubsection{Scene Inclusion Criteria}
The p-shift taxonomy is only structurally meaningful in scenes with sufficient participants. In a three-person scene, the set of possible shifts is mechanically constrained: AB-CA and AB-CD become near-degenerate, since few characters remain outside the active dyad \citep{stivers2021}. We therefore restrict the primary analysis to scenes with \textbf{four or more unique speakers} and at least one male and one female character. This ensures that all five p-shift types are structurally available and that the comparison between homophilous and heterophilous turns is not confounded by scene size artifacts.
This filter yields $N = 27$ scenes and $423$ p-shift events (215 heterophilous, 208 homophilous initial turns). Average speakers per scene: $4.41$ (SD $= 0.80$); gender composition averaged $64\%$ male and $36\%$ female, reflecting the broader male-skewed character distribution documented in Study 1 and in prior corpus-level analyses \citep{bamman2024,leontyeva2024}.
\subsubsection{Statistical Analysis}
We first test overall independence between initial turn type (heterophilous vs.\ homophilous) and p-shift distribution using Pearson's chi-squared test. We then conduct follow-up two-proportion $z$-tests for each p-shift type individually, adjusting for multiple comparisons using the Benjamini-Hochberg false discovery rate (FDR) correction across the five tests. We report FDR-adjusted $p$-values ($p_{\text{BH}}$) and proportions with 95\% Wilson confidence intervals.
\subsection{Results}
\subsubsection{Primary Analysis: 4+ Speaker Scenes}
The overall chi-squared test indicates a significant association between initial turn homophily and p-shift distribution ($\chi^2(4) = 14.16$, $p = .007$). The pattern is consistent with a key structural distinction: heterophilous turns more frequently produce direct reciprocation, while homophilous turns more frequently diffuse the conversation to other participants \citep{gibson2003,hu2021}.
Table~\ref{tab:study3_pshifts} reports proportions and follow-up test results for each shift type. Two shifts survive FDR correction at $\alpha = .05$:
\begin{itemize}
    \item \textbf{AB-BA (reciprocation)} is significantly more likely following heterophilous turns than homophilous turns (66.0\% vs.\ 50.5\%; $z = -3.25$, $p_{\text{BH}} = .006$). Cross-gender exchanges tend to produce tight, dyadic back-and-forth.
    \item \textbf{AB-CA (third-party entry)} is significantly more likely following homophilous turns (15.9\% vs.\ 7.9\%; $z = 2.53$, $p_{\text{BH}} = .028$). When same-gender characters converse, new participants are more likely to enter and address the original speaker.
\end{itemize}
AB-AC (speaker readdress) shows a trend in the same direction — homophilous turns more likely to redirect to a third party (11.1\% vs.\ 5.6\%) — but does not survive correction ($p_{\text{BH}} = .068$). AB-AB (continuation) and AB-CD (parallel shift) show no significant differences.
\begin{table}[t]
\centering
\caption{Participation shift proportions by initial turn homophily (4+ speaker scenes, $N = 423$ events). $p_\text{BH}$: Benjamini-Hochberg adjusted $p$-value. Significant results ($p_\text{BH} < .05$) are in bold.}
\label{tab:study3_pshifts}
\begin{tabular}{@{}lrrrrrr@{}}
\toprule
\textbf{P-Shift} & \textbf{Hetero.\ } & \textbf{Homo.\ } & \textbf{Hetero.\ Prop.} & \textbf{Homo.\ Prop.} & $z$ & $p_\text{BH}$ \\ \midrule
AB-AB & 30 & 31 & 0.140 & 0.149 & 0.28 & .781 \\
AB-AC & 12 & 23 & 0.056 & 0.111 & 2.04 & .068 \\
\textbf{AB-BA} & \textbf{142} & \textbf{105} & \textbf{0.660} & \textbf{0.505} & $-3.25$ & \textbf{.006} \\
\textbf{AB-CA} & \textbf{17} & \textbf{33} & \textbf{0.079} & \textbf{0.159} & \textbf{2.53} & \textbf{.028} \\
AB-CD & 14 & 16 & 0.065 & 0.077 & 0.47 & .781 \\ \bottomrule
\end{tabular}
\end{table}
\subsubsection{Robustness Check: 3+ Speaker Scenes}
To assess whether these patterns generalize beyond strict multi-party scenes, we replicate the analysis on all scenes with three or more unique speakers and at least one character of each gender ($N = 2{,}096$ p-shift events; 1,142 heterophilous, 954 homophilous). The overall association remains significant ($\chi^2(4) = 25.61$, $p < .001$).
The directional pattern is consistent with the primary analysis. AB-BA reciprocation is again more frequent following heterophilous turns (71.3\% vs.\ 60.9\%; $p_{\text{BH}} < .001$). AB-AB continuation is more frequent following homophilous turns (26.3\% vs.\ 20.0\%; $p_{\text{BH}} = .001$) — a result that was non-significant in the 4+ condition, likely because self-continuation is structurally suppressed when more participants are available. The AB-CA and AB-AC diffusion effects trend in the same direction but do not survive correction ($p_{\text{BH}} = .067$ and $.102$, respectively), consistent with reduced interpretability when structural constraints on 3-person scenes are present \citep{stivers2021}.
These results confirm that the core AB-BA finding is not a small-sample artifact of the 4+ filter. The 3+ condition inflates statistical power but introduces the confound that third-party p-shifts are near-impossible in the majority of scenes, which attenuates the diffusion effects that are the primary theoretically motivated contrasts.
\subsection{Discussion}
Study 3 suggests that the gender composition of a speaker addressee pair is associated with the addressee pattern in the subsequent conversational turn. Cross-gender (heterophilous) turns are more likely to produce direct reciprocation: the addressed character turns back and responds in kind, keeping the exchange closed and dyadic \citep{sacks1974,stivers2021}. Same-gender (homophilous) turns are more likely to draw in third parties, either through speaker readdress (AB-AC) or unsolicited entry from a bystander (AB-CA) \citep{gibson2003,hu2021}.
Study 1 shows that male characters are disproportionately the targets of speech 
at the aggregate level. Study 2 suggests that cross-gender dialogue is 
directionally symmetric, within-conversation imbalance is unlikely to be the 
driver. Study 3 suggests a mechanism: homophilous male-to-male exchanges are conversation-spreading — they invite participation from others, extending the network of interactional reach. Heterophilous exchanges, by contrast, tend to close off: they resolve quickly into dyadic reciprocations rather than drawing in new voices. The gendered interaction asymmetry documented in Study 1 is thus not merely a product of who initiates speech, but of the downstream structural consequences of different gender pairings \citep{ibarra1992,mcpherson2001}.
These results are consistent with theoretical accounts of conversational homophily in multi-party interaction~\citep{mcpherson2001,khanam2022}, and extend them from static network membership to dynamic turn-by-turn flow. A key limitation is the relatively small number of independent scenes ($N = 27$ in the primary analysis) and a persistent male-skewed gender composition (64\% male on average), which means homophilous turns are predominantly male-to-male. Future work with more balanced corpora would allow cleaner separation of same-gender effects from the specific dynamics of male homophily.
\section{Conclusion}
Taken together, Studies 1--3 provide a layered account of gendered dialogue structure in film: male characters dominate as both speakers and addressees at the corpus level (Study 1); cross-gender interaction is directionally symmetric on average but concentrated within films exhibiting internally consistent directional 
conventions (Study 2); and same-gender interactions diffuse conversational  attention while cross-gender interactions produce tighter dyadic exchanges (Study 3). These findings complement prior content-based approaches \citep{sap2017,martinez2022} by revealing a structural dimension of gender bias that operates through the architecture of conversation itself, independent of what is being said.

Because screenplays are stylized cultural artifacts rather than transcripts of everyday interaction, the extent to which these patterns are unique to film dialogue remains an open empirical question. Future work could apply comparable addressee-based methods to naturalistic multi-party conversations to test whether gendered asymmetries in conversational targeting and participation shifts reflect broader interactional dynamics or conventions specific to scripted narrative.

\section*{Data and Code Availability}
The analysis code, documentation, and example data format for this project are available at \url{https://github.com/samin-khan/you-talkin-to-me}. Research access to derived speaker-addressee annotations and analysis-ready data may be requested by contacting the author at \texttt{samink@stanford.edu}. To support reuse, the repository includes a dummy example dataset with the same column structure expected by the analysis code.

\section*{Acknowledgments}
This research was supported in its early stages by funding from and collaboration with Stanford SPARQ, a center that builds research-driven partnerships with industry leaders and changemakers to address some of the biggest challenges of our time. We thank the researchers and leadership at Stanford SPARQ for their support in the form of funding, annotation, and thought partnership, all of which made this research possible.
        
\bibliographystyle{plainnat}

\clearpage

\appendix
\section{Full List of Films Analyzed}
The dataset for this project consists of 38 film screenplays. The full list of films included in the analysis is provided in Table~\ref{tab:film_list}.
\begin{table}[h!]
\centering
\caption{The 38 Films Included in the Dataset}
\label{tab:film_list}
\begin{tabular}{@{}ll@{}}
\toprule
\textbf{Film Title} & \textbf{Film Title} \\ \midrule
\textit{(500) Days of Summer} (2009) & \textit{Lincoln} (2012) \\
\textit{Bad Lieutenant} (1992) & \textit{The Life of David Gale} (2003) \\
\textit{Beasts of the Southern Wild} (2012) & \textit{Logan} (2017) \\
\textit{Big Fish} (2003) & \textit{Miami Vice} (2006) \\
\textit{The Bounty Hunter} (2010) & \textit{Nine} (2009) \\
\textit{Cherry Falls} (2000) & \textit{The Nines} (2007) \\
\textit{Crime Spree} (2003) & \textit{The Pianist} (2002) \\
\textit{Dances with Wolves} (1990) & \textit{Pretty Woman} (1990) \\
\textit{Easy A} (2010) & \textit{Prom Night} (2008) \\
\textit{Excalibur} (1981) & \textit{Red Riding Hood} (2011) \\
\textit{Field of Dreams} (1989) & \textit{Shallow Grave} (1994) \\
\textit{Flash Gordon} (1980) & \textit{Speed Racer} (2008) \\
\textit{Fright Night} (1985) & \textit{Taking Sides} (2001) \\
\textit{Gandhi} (1982) & \textit{Last Tango in Paris} (1972) \\
\textit{Ghost Rider} (2007) & \textit{Two for the Money} (2005) \\
\textit{The Godfather Part III} (1990) & \textit{Twister} (1996) \\
\textit{Inglourious Basterds} (2009) & \textit{War of the Worlds} (2005) \\
\textit{Legion} (2010) & \textit{Wild Hogs} (2007) \\
\textit{The Perks of Being a Wallflower} (2012) & \textit{Yes Man} (2008) \\
\bottomrule
\end{tabular}
\end{table}
\end{document}